\documentclass[letterpaper,aps,prb,floatfix,twocolumn,showpacs,superscriptaddress,10pt]{revtex4-1}
\usepackage{graphicx,times}
\usepackage{amssymb,amsmath}
\usepackage{multirow}
\usepackage{color}
\usepackage{epstopdf}

\newcommand{\be}{\begin{equation}}
\newcommand{\ee}{\end{equation}}

\newcommand{\bea}{\begin{eqnarray}}
\newcommand{\eea}{\end{eqnarray}}
\newcommand{\bearr}{\begin{eqnarray}}
\newcommand{\eearr}{\end{eqnarray}}

\newcommand{\nn}{\nonumber}
\usepackage{ifthen}
\newlength{\abstractdiagwidth}
\newlength{\specdiagwidth}
\newcommand{\specdiag}[2][]{\parbox[c]{#2\specdiagwidth}{\includegraphics*[width=#2\specdiagwidth]{#1}}}%
\newcommand{\hon}[1]%
{%
\ifthenelse{\equal{#1}{2_1}}{\specdiag[Fig25]{154}}{}%
\ifthenelse{\equal{#1}{2_2}}{\specdiag[Fig26]{154}}{}%
\ifthenelse{\equal{#1}{4_2}}{\specdiag[Fig27]{191.5}}{}%
\ifthenelse{\equal{#1}{4_5}}{\specdiag[Fig28]{191.5}}{}%
\ifthenelse{\equal{#1}{4_6}}{\specdiag[Fig29]{191.5}}{}%
\ifthenelse{\equal{#1}{4_7}}{\specdiag[Fig30]{191.5}}{}%
}%
\begin{document}

\title{Valence Bond Phases in $S=1/2$   Kane-Mele-Heisenberg Model}

\author{ Mohammad H. Zare}
\affiliation{ Department of Physics, Isfahan University of
Technology, Isfahan 84156-83111, Iran}

\author{ Hamid Mosadeq} 
\affiliation{ Department of Physics, Sharekord University, Shahrekord, Iran}

\author{Farhad Shahbazi} \email{shahbazi@cc.iut.ac.ir}
\affiliation{ Department of Physics, Isfahan University of
Technology, Isfahan 84156-83111, Iran}

\author{S. A. Jafari}
\affiliation{Department of Physics, Sharif University of Technology, Tehran 11155-9161, Iran}

\begin{abstract}
 The  phase diagram of Kane-Mele-Heisenberg (KMH) model in classical limit~\cite{zare}, contains  disordered  regions in the
 coupling space, as the result of to  competition among different terms in the Hamiltonian, leading  to frustration in finding a unique ground state.    
In this work we explore  the nature of these phase in the quantum limit, for a $S=1/2$. Employing  exact diagonalization (ED) in $S_z$ and nearest neighbor valence bond (NNVB) bases, bond and plaquette valence bond mean field theories,
We show that the disordered regions are divided into ordered quantum states in the form of {\em plaquette valence bond crystal }(PVBC) and {\em staggered dimerized}  (SD) phases. 

\end{abstract}

\date{\today}

\maketitle

\section{Introduction}
Two-dimensional frustrated spin systems with $S=1/2$ have lately received  massive attentions, due to
their potential  for realizing the  quantum spin liquid (QSL), a magnetically disordered state which respects
all the symmetries of the systems, even at  absolute zero temperature~\cite{balent}.  
The spin model, recently attracted  many interests, is the Heisenberg model with first and second anti-ferromagnetic exchange interaction,
the $J_{1}-J_{2}$ model, in honeycomb lattice. The lowest coordination number ($z=3$) in 2D, being the unique peculiarity of honeycomb,
makes this lattice a promising candidate to host QSL. It is known that the classical $J_{1}-J_{2}$ model do not show any long range ordering
at $T=0$ for $\frac{1}{6}<\frac{J_{2}}{J_{1}}<0.5$, because of high degeneracy in the energy of ground state~\cite{fouet}. However,
thermal fluctuations are able to lower the free energy of some specific spiral states within the ground state manifold~\cite{kawamura},
a phenomenon called thermal order by disorder~\cite{villain}. So far, many efforts have been devoted  to gain insight into the quantum
nature of this disordered region for $S=1/2$ systems.  Some of these works support the existence of QSL~\cite{qsl1,qsl2,qsl3,qsl4,qsl5,qsl6}
for $0.2\lesssim\frac{J_2}{J_1}\lesssim 0.5$, while others suggest a translational broken symmetry state with plaquette valence bond ordering
for $0.2\lesssim\frac{J_2}{J_1}\lesssim 0.35$ which transforms to a nematic staggered dimerised state when  the ratio $\frac{J_{2}}{J_1}$ rises to lay within
$0.35\lesssim\frac{J_2}{J_1}\lesssim 0.5$~\cite{aron2010,mosadeq,pvb2,pvb3,pvb4,pvb5,pvb6,pvb7}.  For $\frac{J_2}{J_1}>0.5$, a long ranged collinear ordered  ground state is proposed\cite{pvb2,pvb7}.      
 
Quick progresses  in the filed of topological insulators (TI)~\cite{Kane2005,Kane20051,top4,top5,top6,top1,top2,top3},
has drawn the attention of the physicists into the study of the effective spin models in the strong coupling limit of  TI models.
Kane-Mele-Hubbard model, is an example of such models which recently has been studied by various
methods~\cite{kmh1,kmh2,kmh3,kmh4,kmh5,kmh6,kmh7,kmh8,kmh9,kmh10,kmh11,kmh12,kmh13,kmh14,
kmh15,kmh16,kmh17,kmh18,kmh19,kmh20}. The strong coupling (large Coulomb interaction) and weak coupling
(small Coulomb interaction) limits of this model are charachterized by anti-ferromagnetic Mott insulator (AFMI) and topological band insulator (TBI) phases, respectively. For intermediate Coulomb interactions and weak spin-orbit coupling  a gapped QSL phase has been proposed for
his model~\cite{kmh10}.   

The strong coupling  limit of  Kane-Mele-Hubbard model is effectively described by a XXZ model, also called
Kane-Mele-Heisenberg({\rm KMH}) model~\cite{kmh1}. 
Classical phase diagram of  KMH model contains  six regions in the coupling space~\cite{zare}. 
In the three regions the model is long-range ordered, {\em  planar N{\'e}el }state in honeycomb plane (phase I), 
{\em commensurate spiral} states in the  plan normal to honeycomb lattice (phase VI) and
{\em collinear} states along perpendicular to honeycomb plane (phase II).  In the  other three regions the system is disordered, the ground state is infinitely degenerate and characterized by a manifold of incommensurate wave-vectors.
These phases  are, {\em planar spiral} (phase III),
{\em vertical spiral} states (phase IV) and {\em non-coplanar states} (phase V). 
Apart from a Schwinger boson and Schwinger fermion study~\cite{vaezi},     
where a chiral spin liquid state is proposed for a narrow region but large values of second  neighbor exchange interaction ($J_{2}$), 
the quantum phase diagram of KMH model has remain unexplored.

Our aim in this work,  is  understanding the nature of the quantum ground state of  $S=1/2$ KMH model
for intermediate values of  $J_2$,  mostly in phases III and IV, where it is classically disordered. 
For this purpose, we use exact diagonalization as well as valence bond and plaquette mean field theories. 

The paper is organized as follows. In Sec.~\ref{model} the KMH model is introduced. The quantum ground state properties  of the classically disordered phases are investigated, using ED for a finite lattice in Sec.~\ref{ED} and  bond operator and plaquette valence bond mean field theories in Sec.~\ref{BO}. Section \ref{conclusion} is devoted to conclusion.  The details of bond operator and plaquette mean field theories  are given in 
appendices~\ref{appbo} and \ref{apppl}, respectively.

\section{Model Hamiltonian}
\label{model}
Kane-Mele-Hubbard model is described by the following Hamiltonain
\be
H=-t\sum_{\langle ij\rangle,\sigma}c_{i\alpha}^\dagger c_{j\alpha}
+i\lambda\sum_{\langle\!\langle i,j\rangle\!\rangle,\alpha\beta}\nu_{ij}\sigma_{\alpha\beta}^z c_{i\alpha}^\dagger c_{j\beta}+\sum_{i}U n_{i\uparrow}n_{i\downarrow},
\label{kane2.eqn}
\ee
in which $\langle\cdots\rangle$ and $\langle\!\langle \cdots\rangle\!\rangle$ denote the nearest and next to nearest neighbor sites
in a honeycomb lattice. First term represents the hopping  between nearest neighbor atoms, while      
the second  term , with $ \nu_{ij}=\pm 1 $ being an anti-symmetric tensor, 
denotes the hopping between the second  neighbors  arising from the spin-orbit coupling. The last term is onsite Hubbard term, 
in which $U>0$ denotes the Coulomb repulsion energy between two electrons within  a single atom.   
In strong coupling limit, where  $U$ is much larger than $t$ and $\lambda$, the model can be effectively described by a $S=1/2$ spin Hamiltonain, namely the Kane-Mele-Heisenberg ( {\rm KMH }) model~\cite{kmh1}      

\begin{equation}
\begin{split}
  H_{\rm KMH}\!\!&=\!\!J_1\sum_{\langle i,j\rangle}{\bf S}_{i}\cdot {\bf S}_{j}
   +J_2\sum_{\langle\!\langle i,j\rangle\!\rangle}{\bf S}_{i}\cdot {\bf S}_{j}\\
   &+g_2\sum_{\langle\!\langle i,j\rangle\!\rangle}(-S_{i}^x~S_{j}^x-S_{i}^y~S_{j}^y+S_{i}^z~S_{j}^z),
\end{split}
\label{kmh}
\end{equation}
in which $J_{1}=4t^2/U-16{t^4}/U^3$, $J_{2}=4{t^4}/U^3$
  and $g_{2}=4{\lambda^2}/U$ 
are the first and second neighbor exchange couplings.

\section{Exact diagonalizaion}
\label{ED}
\begin{figure}[t]
  \begin{center}
    \includegraphics[scale=0.5]{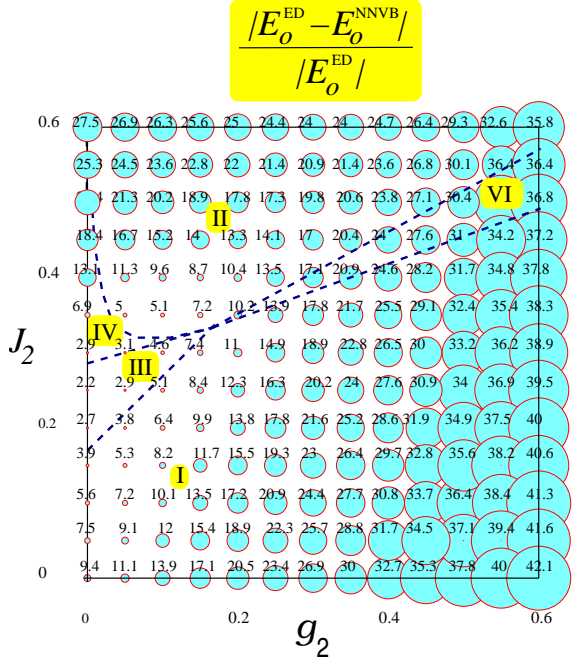}
   \end{center}
   \caption{(Color online) Relative  difference between the ground-state energy obtained by diagonalization in NNVB basis ($E_{0}^{\rm NNVB}$) and  $S_{z}$ basis ($E_{0}^{\rm ED}$)  in coupling space $g_2- J_2$, for $N=24$ lattice points and $J_{1}=1$. Dashed lines display the phase boundaries of the classical KMH model.  The radii of the circles  are proportional to the  relative error, represented in percentage. }
  \label{error}
\end{figure}
\begin{figure}[t]
  \begin{center}
    \includegraphics[scale=0.55]{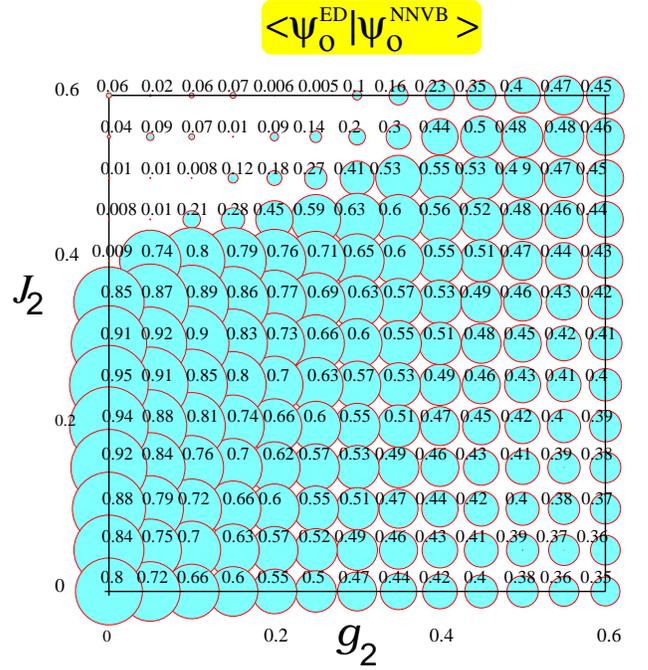}
   \end{center}
   \caption{(Color online)  Overlap between normalized ground state wave functions,  obtained by diagonalization in  NNVB basis ($\psi_{0}^{\rm NNVB}$) and  $S_{z}$ basis ($\psi_{0}^{\rm ED}$) in $g_2-J_2 $, for $N=24$ lattice points and $J_{1}=1$. The radii of the circles are proportional to the magnitude of the overlaps.}
  \label{overlap}
\end{figure}

To gain insight into the fate  of the classically disordered region of ${\rm KMH}$ model in the quantum limit, we employ the exact diagonalization
method in both $S_z$ and nearest neighbor  valence bond (NNVB) bases. NNVB, a basis composed of the products of 
nearest neighbor singlet paris of $S=1/2$ spins, provides a natural framework for characterising the features of the  disordered quantum ground states.
The spin disordered states such as resonating valence bond (RVB) spin liquid and plaquette valence bond crystal  (PVBC) receive most of their components from the the Hilbert space spanning only by  NNVB basis. Therefore, comparing the results of ED within $S_z$ with those  obtained by NNVB basis, would be a guideline to learn about the nature of the ground state in classically degenerated phases III and IV.  

Let us expand the ground state wave function in terms of NNVB states as
\be
   |\psi_0\rangle = \sum_\alpha a(c_\alpha) |c_\alpha\rangle,
   \label{NNVBexp.eqn}
\ee
 where $|c_\alpha\rangle$ denotes all possible configurations $\alpha$ of NNVBs:
\be
   |c_\alpha \rangle = \prod_{(i,j)\in \alpha} (i_\uparrow j_\downarrow-i_\downarrow j_\uparrow).
\ee
First, we have to enumerate the basis $|c_\alpha\rangle$
to construct a numerical representation
of the Hamiltonian matrix in this basis. To determine the basis, the exact Pfaffian
representation of the RVB wave function is employed~\cite{emrrvb}. In this method
one expresses the RVB wave function as the Pfaffian of an antisymmetric matrix
whose dimension is equal to the number of the lattice points. The dimension of Hilbert space corresponding to NNVB
basis is much smaller than the one for whole $S_z=0$ basis, so that the Hamiltonian
matrix can be fully diagonalized with standard library routines.
Note that since the NNVB components ($|c_\alpha\rangle$) are not orthonormal,
one needs to solve the generalized eigen-value problem
$$
   \det [{\cal H}-E{\cal O}] =0,
$$
where ${\cal O}=\langle c_{\beta}|c_{\alpha}\rangle$ denotes the overlap matrix between different NNVB configurations.

We begin with calculation of  relative error in ground state energy between exact and  NNVB basis, $(E_{0}^{\rm NNVB}-E_{0}^{\rm ED})/{E_{0}^{\rm ED}}$ and also the overlap of the corresponding ground state wave functions. From now on we set $J_{1}=1$.
Figs.\ref{error} and \ref{overlap} show the relative errors (in percent) and the overlapping of the ground state wave functions, respectively,   for a system consisting of  $N=24$ lattice points. Relative errors and wave function overlaps indicate that 
the best match between the ground states, obtained by  the two bases, occurs mostly in classically disordered Phase.III and also large part of the phase. IV.  

\begin{figure}[t]
\begin{center}
\includegraphics[scale=0.5]{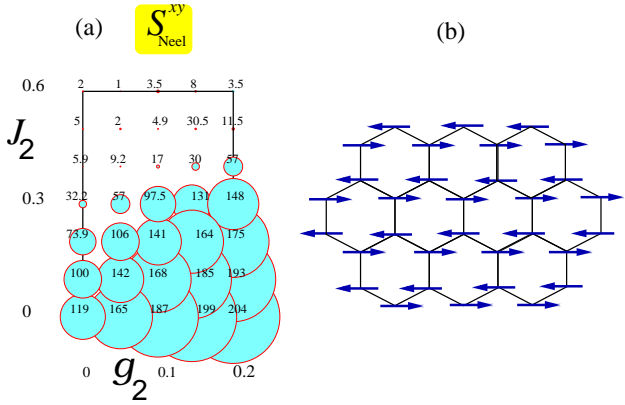}
\caption{(Color online)   Structure function of planar N{\'e}el, calculated by exact diagonalization in $S_z$ basis  for $N=24$ lattice points. The radii of the circles are proportional to the magnitude of the structure function (magnified by the factor 1000) for each point in the coupling space. (b) Schematic representation of N{\'e}el-$xy$ state, proposed for phase.I. }
 \label{Neel}
 \label{Neel}
\end{center}
\end{figure}

\begin{figure}[t]
\centering
\includegraphics[scale=0.3]{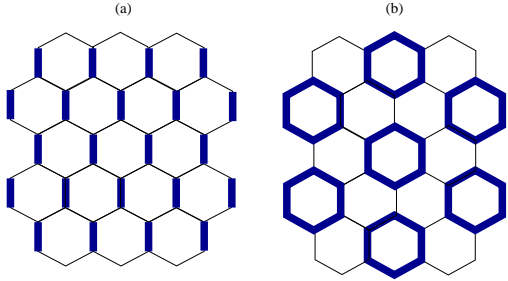}
\caption{ (Color online) Schematic representaion of (a) Staggered dimerized (SD), and  (b)  Plaquette valence bond crystal (PVBC).}
\label{vb}
\end{figure}
\begin{figure}[t]
\begin{center}
\includegraphics[scale=0.5]{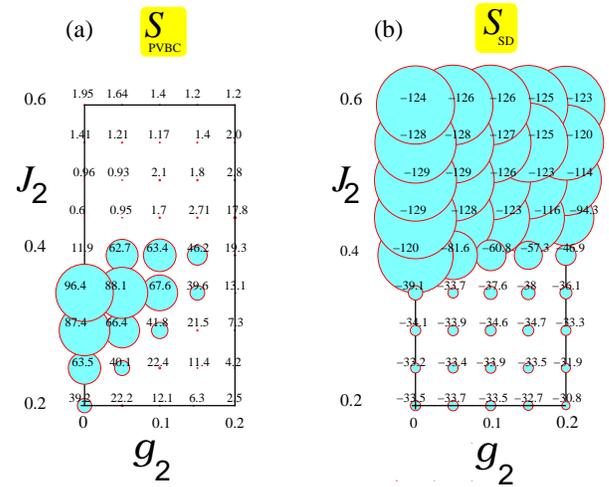}
\caption{(Color online) (a) Plaquette valence bond crystal, and (b) Staggered dimerized  structure functions, calculated by exact diagonalization in $S_z$ basis  for $N=24$ lattice points. The radii of the circles are proportional to the magnitude of the structure function (magnified by the factor 1000) calculated for  each point in the coupling space. 
 \label{SFps}}
\end{center}
\end{figure}

\begin{figure}[t]
\begin{center}
\includegraphics[scale=0.55]{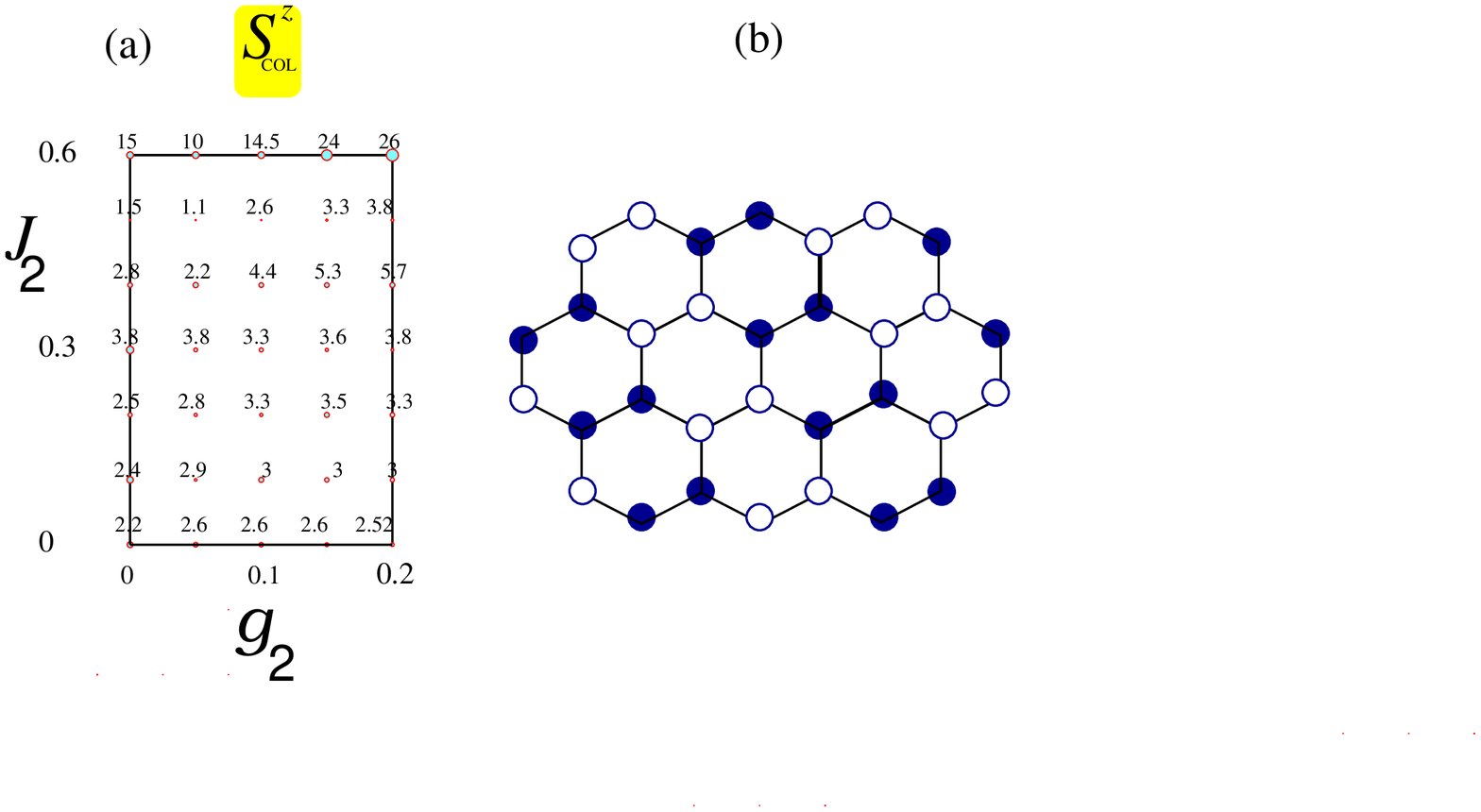}
\caption{(Color online) (a)  Structure function of the collinear  state along $z$-axis, calculated by exact diagonalization in $S_z$ basis  for $N=24$ lattice points. The radii of the circles are proportional to the magnitude of the structure function (magnified by the factor 1000) for  each point in the coupling space. (b) Schematic representation of collinear-$z$ state, proposed for phase II in classical limit. Black and white circles denote the up and down spins, respectively.  
 \label{col}}
\end{center}
\end{figure}

\begin{figure}[t]
\centering
\includegraphics[scale=0.6]{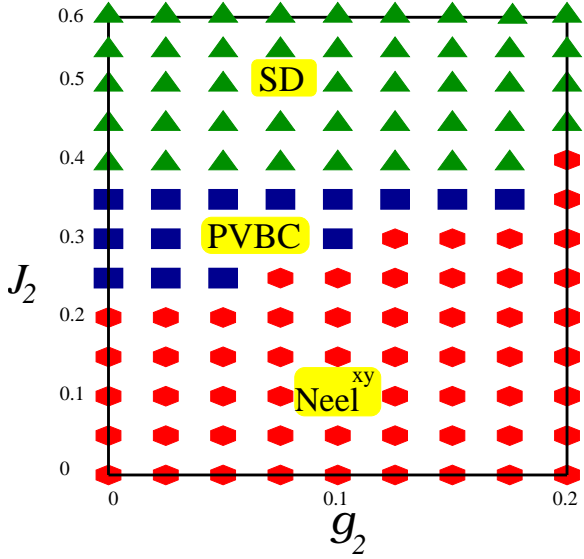}
\caption{(Color online) Quantum phase diagram obtained from exact diagonalization for a finite lattice with $N=24$.}
\label{phase-diagram}
\end{figure}

Now we proceed to inspect the possible orderings in the coupling space by defining appropriate structure functions. Since the spin-orbit coupling is small for real materials, we limit ourselves  to $0 < g_{2} < 0.2 $ and $0 < J_{2} < 0.6$. 
For small values of $J_{2}$, the classical ground state is planar N{\'e}el state. To investigate the region in coupling space where this ordering is extended, we calculate a
structure function corresponding to it in terms of spin-spin correlation functions as

\bea
 S_{\rm Neel}^{xy}=\frac{1}{N^2}&(\sum_{i,j\in A {\rm or} B}\langle S_{i}^{x} S_{j}^{x}+S_{i}^{y} S_{j}^{y}\rangle - \nn\\
 &\sum_{i\in A,j\in B}\langle S_{i}^{x} S_{j}^{x} + S_{i}^{y}S_{j}^{y} \rangle),
\eea
in which $N$ is the number of lattice points and  $A,B$ denote the two sublattices of honeycomb.   

The obtained structure function for N{\'e}el-$xy$ is depicted in Fig~\ref{Neel}, indicating that the N{\'e}el ordering in honeycomb plane extends  to $J_{2}\sim 0.2$ for $g_{2}=0$ and  stretches up to  $J_{2}\sim 0.3$ as $g_{2}$ tends to $0.2$.  

Now we seek the features of the disordered quantum ground state, where N{\'e}el ordering vanishes, and see whether they break any symmetries of the lattice. 
The proposed $SU(2)$ symmetric ground states,  breaking  the symmetries of honeycomb lattice, are the staggered dimerized (SD) or nematic valence bond solid,  which breaks the $C_3$ rotational symmetry  and plaquette valence bond crystal (PVBC) which breaks the translational symmetry of the honeycomb lattice (Fig.\ref{vb}).   The structure functions for SD and PVBC can be defined in terms of dimer-dimer correlations as  
\bea
  S_\lambda=  \frac{1}{N_b}\sum_{\alpha'} \varepsilon_\lambda(\alpha')~C(\alpha,\alpha'),
\eea
where $N_b$ denotes the number of bonds and $C(\alpha,\alpha')$ is the dimer-dimer correlation given by
\be
    C(\alpha,\alpha')=
   4(\langle(\textbf{S}_i.\textbf{S}_j)(\textbf{S}_k.\textbf{S}_l)\rangle-
   {\langle (\textbf{S}_i.\textbf{S}_j)\rangle}^2),
   \label{ddcorr.eqn}
\ee
where $\alpha'=(k,l)$, and $\alpha=(i,j)$ denotes  the reference bond relative to which the correlations
are calculated. $\varepsilon_\lambda(\alpha')$ is the phase factor, appropriately
defined for each of the two states $\lambda\equiv$SD, PVBC~\cite{mambrini}.

The two structure functions, calculated exactly in $S_z$ basis for $N=24$,  are represented in Fig.~\ref{SFps}, where the 
radii of the circles denote the strength of  aforementioned orderings  for each set of couplings $(g_2,J_2)$. Fig.~\ref{SFps}-a shows that in the most part of  phases III and  IV, where the ground state is well described by NNVB basis, the PVBC structure function is remarkably large, while for  $J_2\gtrsim 0.4$, it falls down abruptly. On the other hand, Fig~\ref{SFps}-b shows the sudden growth of SD structure function for $J_{2}\gtrsim 0.4$, the indication of  first order phase transition between PVBC and SD states.  As can also be seen from this figure,  for the range of coupling under study, the  SD ordering is well developed inside the phase II, for which a collinear ordering perpendicular to the honeycomb plane is found in classical limit. The structure function corresponding to collinear-$z$ ordering, for which a possible configuration is depicted  in Fig~\ref{col}-b, can be defined as 
\begin{equation}
S_{{\rm COL}}^{z}=\frac{1}{N^2}\sum_{i,j}e^{i[{\bf q}.({\bf r}_i-{\bf r}_j)]}\langle S_{i}^{z} S_{j}^{z}\rangle,
\end{equation}
in which ${\bf q}=(\pi,\pi/\sqrt{3})$,  ${\bf r}_{i}$ denotes the translational  vector of triangular Bravais lattice and the  unit cell is chosen in such a way to contain  two parallel spins. 
Fig~\ref{col}-a displays the values of $S_{\rm COL}^{z}$ obtained from ED calculation.  The magnitudes of this structure function, being very small compare to the ones corresponding to SD ordering,   verify  the alternation  of SD  ordering instead of  collinear-$z$ state in phase II, at least for  $g_{2} < 0.2$. 

The results of this section is summarized in a finite lattice quantum phase diagram, represented in Fig.~\ref{phase-diagram}.

\section{Bond Operator Method}
\label{BO}  

Inspired by  ED calculation on the finite system,  in this section we employ bond operator as well as plaquette operator  mean-field theories to investigate the regions of the stability of PVBC and SD phases and  transition between them, for the infinite lattice. 

The bond operator formalism is introduced by Chubokov~\cite{chubukov} and Sachdev and Bhatt~\cite{sachdev}, for describing the disordered phases of a frustrated spin Hamiltonian. 
In this formalism, a couple of $S=1/2$ spin operators  belonging  to a bond are represented in terms of  the components of   their summation, with a Hilbert space  
consisting of one singlet $|s  \rangle$ and three triplet states $|t_x\rangle$, $|t_y\rangle$ and $|t_z\rangle$.
Introducing, the singlet and triplet creation operators  out of vacuum $|0\rangle$
\begin{eqnarray}
 |s  \rangle&=&s^{\dag}|0\rangle=\frac{1}{\sqrt{2}}(|\uparrow\downarrow\rangle-|\downarrow\uparrow\rangle)\nonumber\\ 
 |t_x\rangle&=&{t_x}^{\dag}|0\rangle=\frac{-1}{\sqrt{2}}(|\uparrow\uparrow\rangle-|\downarrow\downarrow\rangle)\nonumber \\ 
 |t_y\rangle&=&{t_y}^{\dag}|0\rangle=\frac{i}{\sqrt{2}}(|\uparrow\uparrow\rangle+|\downarrow\downarrow\rangle) \nonumber \\ 
 |t_z\rangle&=&{t_z}^{\dag}|0\rangle=\frac{1}{\sqrt{2}}(|\uparrow\downarrow\rangle+|\downarrow\uparrow\rangle),\nonumber\\
\label{bo-rep}
\end{eqnarray}
one can  express a spin residing  on site $n$, in terms of these basis states as ${\bf S}_n=\sum_{\mu,\nu} |\mu \rangle\langle \mu |{\bf S}_n|\nu\rangle \langle \nu| $. Here $|\mu\rangle$ and $|\nu\rangle$ can be each of the above four states.
Evaluation of the  matrix elements $\langle \mu |{\bf S}_n|\nu\rangle$,  gives rise to the 
 representation of the spin operator in terms of the bosonic bond operators
\begin{equation}
\label{bo-rep}
S_{n}^{\alpha}= \frac{{(-1)}^n}{2}(s^{\dag}t_\alpha+t_{\alpha}^{\dag}s)-\frac{i}{2}{\epsilon}_{\alpha \beta \gamma} t_{\beta}^{\dag}{t_{\gamma}},
\end{equation}
where $\alpha$, $\beta$ and $\gamma$ stand for $x$,$y$ and $z$ and $\epsilon$ is the totally  anti-symmetric tensor.
Moreover, the fact  that each bond is either in a singlet or triplet state,  leads to the following constraint
\begin{equation}
 s^{\dag}s+\sum_{\alpha}{t_{\alpha}}^{\dag}{t_{\alpha}}=1
\label{const}
\end{equation}
Now, considering  a SD configuration illustrated in Fig.~\ref{vb}-a, the spin Hamiltonian \ref{kmh} can be decomposed into the inter and intra bond
terms given by Eq.~\ref{sd}. Using the spin representations \ref{bo-rep}, we achieve a bosonic Hamiltonian in terms of  singlet and triplet
operators, in which all the singlets are considered to be condensed. Then, keeping only the quadratic triplet terms, as an approximation, enables us
to diagonalize the resulting Hamiltonian by the use of  Bogoliubov transformations. Finally, minimization of  the total energy subjected to the
constraint \ref{const}, provides us  with a set of self consisted equations. Numerical solution of these equations gives the energy of
corresponding dimerized configuration. The details of the derivation of self-consisted equations are given in appendix \ref{appbo}.    

In order to find the energy of a plaquette ordered state, we rewrite the spin Hamiltonian \ref{kmh} in terms of  the {\em plaquette operators}
defining  based on the eigenstates of  KMH Hamiltonain for a single hexagon. 
In the absence of Kane-Mele term, {\em i.e} $g_{2}=0$,  The commutation relation $[H,{\bf S}^2]=0$, enables us to label  
each eigenstate of such a  Hamiltonian by the eigenvalues of ${\bf S}^2$ operator.
The ground state is then found to be a spin singlet,  invariant under rotation by $60^0$,  up to $J_{2}/J_{1}=$0.5.
This ground state is predominantly expressed by the symmetric combination of  two Kekule structures,  implying  that the ground state 
of $J_{1}-J_{2}$ within a hexagon is  s-wave singlet, in contrast to the f-wave singlet (the anti-symmetric superposition of two Kekule structures)
proposed in~\cite{pvb5}.  
The first excited states are also found to be  triplet for $0<J_{2}/J_{1}<0.25$ and  replaced by a f-wave singlet state  for  $J_{2}/J_{1}>0.25$.

\begin{figure}[t]
\begin{center}
\includegraphics[scale=0.4]{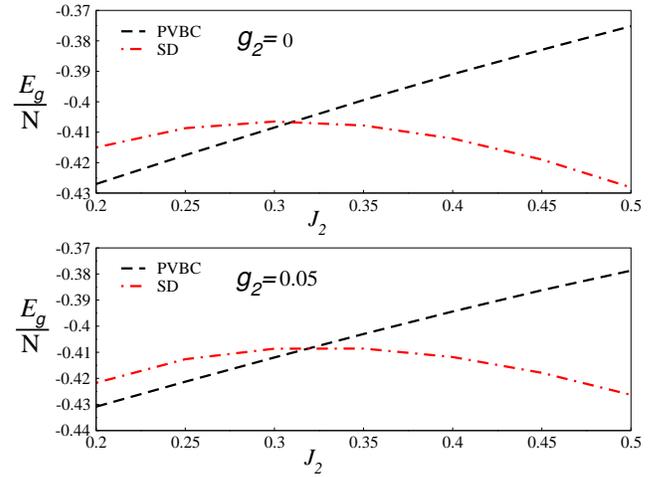}
\caption{(Color online) The ground state energy per spin obtained  from bond operator method for SD state (dotted-dashed line) and from plaquette operator method for PVBC state ( dashed line)  versus $J_2$ for  (top) $g_2=0.0$ and (bottom) $g_2=0.05$.}
\end{center}
\label{e-bopl}
\end{figure}

Now, we proceed to represent the spin operators in terms of the eigenstates of the $J_{1}-J_{2}$ Hamiltonian within a hexagon. 
The spin operators connect the s-wave ground state singlet only to the triplet excited states, hence, we need to seek the ground state of
full Hamiltonian  in subspace of the Hilbert space consisting of s-wave singlet and triplet states. Therefore the relevant matrix elements are 
\begin{equation}
 a_{n,m}=\langle s_{1} | S_{n\alpha }|t_{m\alpha}\rangle,
\end{equation}
in which $|s_{1} \rangle$ and $|t_{m\alpha}\rangle$  are the s-wave singlet and triplet excited stats, respectively.  
These  states can be represented in terms of creation and annihilation operators, as
\begin{eqnarray} 
 s_{1}^{\dag}| 0\rangle&=&|S_{tot}=0;S_z=0\rangle\nonumber\\ 
 t_{1x}^{\dag}|0\rangle&=&\frac{-1}{\sqrt{2}}(|S_{tot}=1;S_z=1\rangle-|S_{tot}=1;S_z=-1\rangle) \nonumber\\ 
 t_{1y}^{\dag}|0\rangle&=&\frac{ i}{\sqrt{2}}(|S_{tot}=1;S_z=1\rangle+|S_{tot}=1;S_z=-1\rangle) \nonumber \\ 
 t_{1z}^{\dag}|0\rangle&=&|S_{tot}=1;S_z=0\rangle.\nonumber\\
\end{eqnarray}
We can  represent the  spin at site $n$ as
\begin{equation}
 S_{n\alpha }= \sum_{m}{a}_{n,m}(s_{1}^\dag t_{m\alpha}+{t^\dag}_{m\alpha}s_{1}).
\label{SRP}
\end{equation}
 Restricting  to the reduced Hilbert space, requires the following constraint 
\begin{equation}
 s_{1}^\dag s_{1} +\sum_{m,\alpha} {{t^\dag}_{m\alpha}}{{t}_{m\alpha}}=1
 \label{const2}
\end{equation}
The procedure similar to bond operator method leads to a set of self-consistent equations from which we can calculate the ground state
energy corresponding to plaquette ordered state. For more details we refer the reader to appendix \ref{apppl}. 

Fig.~\ref{e-bopl}, shows two  plots of energy per spin for SD and PVBC states as a function of $J_{2}$ for $g_{2}=0$ (top panle)
and $g_{2}=0.05$ (bottom panel).
Both plots illustrates the crossing of PVBC and SD energies upon  as $J_{2}$ is increased. For $g_{2}=0$, the transition point between PBVC to SD
is at $J_{2}\sim 0.31$ and increases a little by rising  the value of $g_{2}$. The crossing of the two energies indicates that the transition is
first order. 

\begin{figure}[t]
\begin{center}
\includegraphics[scale=0.6,angle=0]{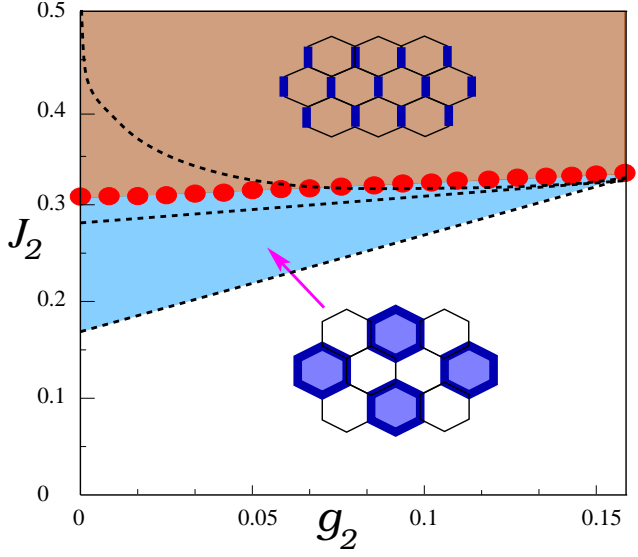}
\caption{(Color online) Phase diagram of KMH model obtained by bond operator and plaquette operator methods.}
\label{pd-bopl}
\end{center}
\end{figure}

As the final result the bond and plaquette operator phase diagram of KMH model are represented in Fig.~\ref{pd-bopl}, showing that for
$J_{2}\lesssim 0.3 $, i.e the classical phase III and the lower part of the classical phase IV, the ground state is a PVBC while for upper part of phase IV and also
inside the the classical phase II the ground state is described by an SD state, in qualitative agreement with ED results for the finite lattice.     

\section{Conclusion}
\label{conclusion}
In summary, we explored the quantum phase diagram of the $S=1/2$ KMH model,  using of exact diagonalizaion for a finite lattice and
a bond operator and plaquette operator methods for infinite system size, with the focus on  the  regions of coupling space with high classically degeneracy. 
Here, we found that the N{\'e}el, PVBC and SD orderings found for  $J_{1}-J_{2}$ Heisenberg model, adiabatically continues to the phase space of  KMH model. The effect of spin-orbit term $g_{2}$, which reduces the $O_{3}$ symmetry of Heisenberg model to $O_2$ for KMH, is converting the isotropic N{\'e}el ordered (for $0 < J_{2} \lesssim 0.2$) state to a planar N{\'e}el ordering in honeycomb plane.  
Moreover, the PVBC  ordered state which is found to be the ground state of $J_{1}-J_{2}$ model, for $0.2 \lesssim J_{2}/J_{1} \lesssim 0.35$,
is adiabaticically continued into the classical phase III  and lower part of phase IV. For $0.35 \lesssim J_{2}/J_{1} \lesssim 0.5$,
the SD ordering obtained for isotropic  model extends toward upper part of phase IV and also into classically ordered phase II for $g_{2}<0.2$.
Our work highlights the significance of quantum fluctuations for $S=1/2$ KMH model, in melting down the classically ordered state  
into  purely quantum ground states. 

\appendix

\section{ Self-consistent equation of bond operator mean field theory}
\label{appbo}

The spin Hamiltonian \ref{kmh} for an SD configuration can be rewritten as  
\begin{equation}
\begin{split}
 H&=J_1{\sum}_{\langle ij \rangle \in {\rm bond}} {\bf S}_i.{\bf S}_j+ (J_2-g_2){\sum}_{\langle ij \rangle \in{\rm bond}}{\bf S}_i.{\bf S}_j\\
 &+2g_2{\sum}_{\langle ij \rangle \in {\rm bond}}{S}_i^{z}{S}_j^{z}+J_1{\sum}_{\langle ij \rangle \notin {\rm bond}} {\bf S}_i.{\bf S}_j\\
 &+(J_2-g_2){\sum}_{\langle ij \rangle \notin {\rm bond}}{\bf S}_i.{\bf S}_j+2g_2{\sum}_{\langle ij \rangle \notin {\rm bond}}{S}_i^{z}{S}_j^{z}.
 \label{sd}
\end{split}
\end{equation}
Inserting the  spin representations  \ref{bo-rep} into this Hamiltonian, 
assuming that all the singlets are condensed (this means replacing  $s$ and $s^{\dagger}$ with the c-number $\bar{s}$),
 keeping only the quadratic terms, incorporating the constraint \ref{const} by a Lagrange multiplier $\mu$,  
 and finally the Fourier transformation, we obtain the following quadratic Hamiltonian 
in terms of the momentum space triplet operators 
\begin{eqnarray}
 && {H^{[2]}}_{BO}= -N_{b}\frac{3}{4}J_1 {\overline{s}}^2 - N_{b}\mu {\overline{s}}^2 +N_{b}\mu + \nonumber \\ 
&&  {\sum}_{k>0}[ (G_k+F_k^{1})({t^\dag}_{k,{x}}{t_{k,{x}}}+{t^\dag}_{-k,{x}}{t_{-k,{x}}})+\nonumber\\ 
&&  (G_k+F_k^{1})({t^\dag}_{k,{y}}{t_{k,{y}}}+{t^\dag}_{-k,{y}}{t_{-k,{y}}})+\nonumber\\
&&  (G_k+F_k^{2})({t^\dag}_{k,{z}}{t_{k,{z}}}+{t^\dag}_{-k,{z}}{t_{-k,{z}}})+\nonumber\\
&&  (F_k+F_k^{1})({t^\dag}_{k,{x}}{{t^\dag}_{-k,{x}}}+{t}_{k,{x}}{t_{-k,{x}}})+\nonumber\\
&&  (F_k+F_k^{1})({t^\dag}_{k,{y}}{{t^\dag}_{-k,{y}}}+{t}_{k,{y}}{t_{-k,{y}}})+\nonumber\\
&&  (F_k+F_k^{2})({t^\dag}_{k,{z}}{{t^\dag}_{-k,{z}}}+{t}_{k,{z}}{t_{-k,{z}}})],
\label{h2bo}
\end{eqnarray}
 where $N_b$ is the number of bonds and 
\begin{eqnarray}
 &&G_k=\frac{J_1}{4}-\mu-\frac{{{\overline{s}}^2}}{4}J_1({\epsilon}_k+{\epsilon}_{-k})
      +\frac{{{\overline{s}}^2}}{4}J_2({\eta}_k+{\eta}_{-k})\nonumber\\
 &&F_k= -\frac{{{\overline{s}}^2}}{4}J_1({\epsilon}_k+{\epsilon}_{-k})
       +\frac{{{\overline{s}}^2}}{4}J_2({\eta}_k+{\eta}_{-k}) \nonumber\\
 &&F_k^{1}= -\frac{{{\overline{s}}^2}}{4}g_2({\eta}_k+{\eta}_{-k})   \nonumber\\
 &&F_k^{2}= \frac{{{\overline{s}}^2}}{4}g_2({\eta}_k+{\eta}_{-k}).
\end{eqnarray}
In the above relations ${\epsilon}_k$ and $\eta_k$ are defined as
\begin{eqnarray}
   \epsilon_k &=&e^{-ik_b}+e^{-i(k_b+k_a)}\nonumber\\
    \eta_k &=& 2 \left[\cos(k_a)+\cos(k_b)+\cos(k_a+k_b) \right]
\end{eqnarray}
Using appropriate  Bogoliubov  transformations, the  Hamiltonian \ref{h2bo} can be  diagonalized as
\begin{eqnarray}
&&{H^{[2]}}_{BO}= N_{b}(-\frac{3}{4}J_1 {\overline{s}}^2 - \mu {\overline{s}}^2 +\mu) + \nn \\
&& {\sum}_{k>0}({\omega}_{k,x}+{\omega}_{k,y}+{\omega}_{k,z}-3G_k-2F_k^{1}-F_k^{2})+ \nn \\ 
&&  {\sum}_{k>0}{\omega}_{k,x}({\gamma}_{k,x}^{\dag}{{\gamma}}_{k,x}+
   {\gamma}_{-k,x}^{\dag}{{\gamma}}_{-k,x})+\nn\\
&& {\sum}_{k>0}{\omega}_{k,y}({\gamma}_{k,y}^{\dag}{{\gamma}}_{k,y}+
   {\gamma}_{-k,y}^{\dag}{{\gamma}}_{-k,y})+ \nn \\
&& {\sum}_{k>0}{\omega}_{k,z}({\gamma}_{k,z}^{\dag}{{\gamma}}_{k,z}+
   {\gamma}_{-k,z}^{\dag}{{\gamma}}_{-k,z})   ,
\end{eqnarray}
in which 
\begin{eqnarray}
&& \omega_{k,x}=\omega_{k,y}=\sqrt{({G_k}+F_k^{1})^2-({F_k}+F_k^{1})^2}\nn\\
&& \omega_{k,z}=\sqrt{({G_k}+F_k^{2})^2-({F_k}+F_k^{2})^2} , 
\end{eqnarray}
are the  triplon dispersions  and
\begin{eqnarray}
 &&  {\epsilon}_{g}=(-\frac{3}{4}J_1 {\overline{s}}^2 - \mu {\overline{s}}^2 +\mu) + \nn\\
&&{\sum}_{k>0}({\omega}_{k,x}+{\omega}_{k,y}+{\omega}_{k,z}-3G_k-2F_k^{1}-F_k^{2}),
\end{eqnarray}
gives the ground state energy per bond. 
The ground state energy depends on the parameters $\mu$ and ${\bar s}$, 
and can be determined self-consistently from  the saddle-point conditions
\bea
 \frac{\partial \epsilon_{g}}{\partial \mu}&=&
  -\overline{s}^2+1-2{\sum}_{k>0}\frac{(G_k+F_k^{1})}{\omega_{k,x}} \nn\\
 & - &{\sum}_{k>0}\frac{(G_k+F_k^{2})}{\omega_{k,z}}+ 3{\sum}_{k>0} 1=0,\nn\\ 
  \frac{\partial \epsilon_{g}}{\partial {\overline{s}}^2}&=&-\frac{3}{4}J_1-\mu\nn\\
   &+&2{\sum}_{k>0}\left(-\frac{J_1}{4}({\epsilon}_k+{\epsilon}_{-k})+\right.
   \left.\frac{(J_2-g_2)}{4}({\eta}_k+{\eta}_{-k})\right)\nn\\
   &\times&(\frac{G_k-F_k}{\omega_{k,x}}-1)\nn\\
  &+&{\sum}_{k>0}\left(-\frac{J_1}{4}({\epsilon}_k+{\epsilon}_{-k})+\frac{(J_2-g_2)}{4}({\eta}_k+{\eta}_{-k})\right)\nn\\
  &\times&(\frac{G_k-F_k}{\omega_{k,z}}-1)=0.
\eea

\section{ Self consistent equations of plaquette mean field theory}
\label{apppl}
Considering the PVBC ordering shown in  Fig.\ref{vb}-b,  the Hamiltonian \ref{kmh} can be rewrited as 
\begin{equation}
\begin{split}
 H&=J_1{\sum}_{\langle ij \rangle \in PL} {\bf S}_i.{\bf S}_j+ (J_2-g_2){\sum}_{\langle ij \rangle \in PL}{\bf S}_i.{\bf S}_j\\
 &+2g_2{\sum}_{\langle ij \rangle \in PL}{S}_i^{z}{S}_j^{z}+J_1{\sum}_{\langle ij \rangle \notin PL} {\bf S}_i.{\bf S}_j\\
 &+(J_2-g_2){\sum}_{\langle ij \rangle \notin PL}{\bf S}_i.{\bf S}_j+2g_2{\sum}_{\langle ij \rangle \notin PL}{S}_i^{z}{S}_j^{z}.
 \label{SPH1}
\end{split}
\end{equation}
The Hamiltonian of a single hexagonal block can be represented in terms of creation and annihilation operators as 
\begin{equation}
 H_{PL}= \sum_{p}\epsilon_{s_{p}} s_{p}^\dag s_{p} + \sum_{m}\epsilon_{t_{m}}{{t^\dag}_{m\alpha}}{{t}_{m\alpha}},
\end{equation}
where $\epsilon_s$ and $\epsilon_{t_{m}}$ are evaluated numerically by diagonalizing the KHM Hamiltonian  in $S_Z=0$ basis in a hexagon.
Reexpressing the Hamiltonian Eq.\ref{SPH1} in these new singlet and triplet operators and, incorporating the constraint \ref{const2}, using  the  Bogoliubov transformations,  and assuming the condensation of singlets, we arrive at the following diagonalized Hamiltonian in ${\bf k}$-space, 
\begin{eqnarray}
&&{H^{[2]}}_{PL}= N_{p}(\overline{s}^2\epsilon_{s_{1}} - \mu \overline{s}^2 +\mu) + \nn\\
&&{\sum}_{k>0}({\omega}_{m,k}^{x}+{\omega}_{m,k}^{y}+{\omega}_{m,k}^{z}-3G_{m,k}-2G_{m,k}^{1}-G_{m,k}^{2})+ \nn \\ 
&&  {\sum}_{k>0}{\omega}_{m,k}^{x}({\gamma}_{m,kx}^{\dag}{{\gamma}}_{m,kx}+
   {\gamma}_{m,-kx}^{\dag}{{\gamma}}_{m,-kx})+\nn\\
&& {\sum}_{k>0}{\omega}_{m,k}^{y}({\gamma}_{m,ky}^{\dag}{{\gamma}}_{m,ky}+
   {\gamma}_{m,-ky}^{\dag}{{\gamma}}_{m,-ky})+\nn  \\
&& {\sum}_{k>0}{\omega}_{m,k}^{z}({\gamma}_{m,kz}^{\dag}{{\gamma}}_{m,kz}+
   {\gamma}_{m,-kz}^{\dag}{{\gamma}}_{m,-kz})   ,                          
\end{eqnarray}
where
\begin{eqnarray}
&&G_{m,k}=\epsilon_{t_m}-\mu+2J_1\overline{s}^2S_{m,k}^{1}+J_2S_{m,k}^{2} \nn\\
&&F_{m,k}=2J_1\overline{s}^2S_{m,k}^{1}+J_2S_{m,k}^{2} \nn\\
&&G_{m,k}^{1}= -g_2S_{m,k}^{2}=-G_{m,k}^{2}\nn\\
&&S_{m,k}^{1}={a}_{2,m}{a}_{5,m}\cos(k_a)+{a}_{3,m}{a}_{6,m}\cos(k_b)+\nn\\
&&{a}_{1,m}{a}_{4,m}\cos(k_a+k_b)\nn\\
&&S_{m,k}^{2}=({a}_{5,m}{a}_{3,m}+{a}_{1,m}{a}_{5,m}+{a}_{2,m}{a}_{6,m}+{a}_{2,m}{a}_{4,m})\nn\\
&&\times\cos(k_a)+({a}_{1,m}{a}_{3,m}+{a}_{3,m}{a}_{5,m}+{a}_{4,m}{a}_{6,m}+{a}_{2,m}{a}_{6,m})\nn\\
&&\times\cos(k_b)+({a}_{1,m}{a}_{3,m}+{a}_{1,m}{a}_{5,m}+{a}_{4,m}{a}_{6,m}+{a}_{2,m}{a}_{4,m})\nn\\
&&\times\cos(k_a+k_b),
\end{eqnarray}
in which
\begin{eqnarray}
&& \omega_{m,k}^{x}=\omega_{m,k}^{y}=\sqrt{({G_{m,k}}+G_{m,k}^{1})^2-({F_{m,k}}+G_{m,k}^{1})^2}\nn\\
&& \omega_{m,k}^{z}=\sqrt{({G_{m,k}}+G_{m,k}^{2})^2-({F_{m,k}}+G_{m,k}^{2})^2} ,
\end{eqnarray}
are the triplon dispersions and
\begin{eqnarray}
\label{epl}
&&   {\epsilon}_{g}=(J_1 {\overline{s}}^2{\epsilon_{s_1}} - \mu {\overline{s}}^2 +\mu) + \nn\\
&& {\sum}_{k>0}({\omega}_{m,k}^{x}+{\omega}_{m,k}^{y}+{\omega}_{m,k}^{z}-3G_{m,k}-2G_{m,k}^{1}-G_{m,k}^{2}),\nn\\
\end{eqnarray}
is the ground state energy per plaquette. Minimization of  \ref{epl} with respect to the chemical potential $\mu$ and condensate density ${\bar s}$, gives rise to the following self-consistent  equations  
\bea
  \frac{\partial \epsilon_{g}}{\partial \mu}&=&
  -\overline{s}^2+1-2{\sum}_{k>0}\frac{(G_{m,kz}+G_{m,kz}^{2})}{\omega_{m,k}^{z}} \nn\\
  &-& 2{\sum}_{k>0}\frac{(G_{m,kx}+G_{m,kx}^{1})}{\omega_{m,k}^{x}}+ 3{\sum}_{k>0} 1=0,\nn\\ 
  \frac{\partial \epsilon_{g}}{\partial {\overline{s}}^2}&=&{\epsilon_s}-\mu \nn\\
  &+&{\sum}_{k>0}\left(({\epsilon_{t,m}^{z}-\mu})
  \frac{J_{1}S_{m,kz}^{1}+(J_2+g_2)S_{m,kz}^{2}}{\omega_{m,k}^{z}}\right)\nn\\
   &+&2{\sum}_{k>0}\left(({\epsilon_{t,m}^{x}-\mu}) \frac{J_{1}S_{m,kx}^{1}+(J_2-g_2)S_{m,kx}^{2}}{\omega_{m,k}^{x}}\right)\nn\\
  &-&2{\sum}_{k>0}\left(J_{1}S_{m,kx}^{1}+(J_2-g_2) S_{m,kx}^{2}\right)\nn\\
  &-&{\sum}_{k>0}\left(J_{1}S_{m,kz}^{1}+(J_2+g_2) S_{m,kz}^{2}\right)=0
\eea
whose solution provides us with the ground state energy of PVBC state.


\begin{thebibliography}{99}
\bibitem{balent} L. Balent, Nature(London) {\bf 464}, 199(2010).
\bibitem{fouet} J. B. Fouet, P. Sindzingre,  and C. Lhuillier, Eur. Phys. J. B {\bf 20}, 241 (2001).
\bibitem{kawamura} S. Okumura, H. Kawamura, T. Okubo, and Y. Motome, J. Phys. Soc. Jpn. {\bf 79}, 114705 (2010).
\bibitem{villain} J. Villain, Z. Phys. B: Condens. Matter {\bf 33}, 31 (1979).
\bibitem{qsl1} F. Wang, Phys. Rev. B \textbf{82}, 024419 (2010).
\bibitem{qsl2} B. K. Clark, D. A. Abanin, and S. L. Sondhi, Phys. Rev. Lett. {\bf 107}, 087204 (2011).
\bibitem{qsl3}   D. C. Cabra, C. A. Lamas, and H. D. Rosales  Phys. Rev. B {\bf 83}, 094506 (2011).
\bibitem{qsl4} Y-M. Lu, and Y. Ran, Phys. Rev. B \textbf{84}, 024420 (2011).
\bibitem{qsl5} H. Zhang, and C. A. Lamas, Phys. Rev. B {\bf 87}, 024415 (2013).
\bibitem{qsl6} X-L. Yu, D-Y Liu, P. Li, and L-J.  Zou, Physica E {\bf 59} 41 (2014).
\bibitem{aron2010} A. Mulder, R. Ganesh, L. Capriotti, and A. Paramekanti, \textbf{81}, 214419 (2010).
\bibitem{mosadeq} H. Mosadeq, F. Shahbazi, and S. A. Jafari, J. Phys.: Condens. Matter, \textbf{23}, 226006 (2011).
\bibitem{pvb2} A. F. Albuquerque, D. Schwandt, B. Hetenyi, S. Capponi, M. Mambirini, and A. M. Lauchli, Phys. Rev. B {\bf 84}, 024406 (2011).
\bibitem{pvb3} J. Reuther, D. A. Abanin, and T. Thomale, Phys. Rev. B {\bf 84}, 014417 (2011).
\bibitem{pvb4} P. H. Y. Li, R. F. Bishop, D. J. J. Farnell,  and C. E. Campbell, J. Phys.: Condens. Matter {\bf 24},  236002 (2012); Phys. Rev. B {\bf 86}, 144404 (2012).
\bibitem{pvb5} R. Ganesh,  S. Nishimoto, and J. van den Brink, Phys. Rev. B {\bf 87}, 054413 (2013).
\bibitem{pvb6} Z. Zhu, D. A. Huse, and S. R. White, Phys. Rev. Lett. {\bf 110}, 127205 (2013).
\bibitem{pvb7} R. F. Bishop, P. H. Y. Li,  and C. E. Campbell, J. Phys.: Condens. Matter {\bf 25} , 306002 (7pp) (2013).
\bibitem{Kane2005} C. L. Kane and E. J. Mele, Phys. Rev. Lett {\bf 95}, 226801 (2005).
\bibitem{Kane20051} C. L. Kane and E. J. Mele, Phys. Rev. Lett. {\bf 95}, 146802 (2005).
\bibitem{top4} B. A. Bernevig and S.-C. Zhang, Phys. Rev. Lett. {\bf 96},106802 (2006).
\bibitem{top5} B. A. Bernevig, T. L. Hughes, and S.-C. Zhang, Science {\bf 314}, 1757 (2006).
\bibitem{top6} M. K\"{o}nig, S. Wiedmann, C. Br\"{u}ne, A. Roth, H. Buhmann, L. W. Molenkamp, X.-L. Qi, and S.-C. Zhan, Science {\bf 318}, 766 (2007).
\bibitem{top1} M. Z. Hasan and C. L. Kane, Rev. Mod. Phys. {\bf 82}, 3045 (2010).
\bibitem{top2} X.-L. Qi and S.-C. Zhang, Rev. Mod. Phys. {\bf 83}, 1057 (2011).
\bibitem{top3} B. A. Bernevig, {\em Topological Insulators and Topological Su-perconductors} (Princeton University Press, Princeton and
Oxford, 2013).
\bibitem{kmh1} S. Rachel and K. Le Hur, Phys. Rev. B {\bf 82}, 075106 (2010).
\bibitem{kmh2} D. Soriano and J. Fern{\' a}ndez-Rossier, Phys. Rev. B {\bf 82},161302 (2010).
\bibitem{kmh3} Y. Yamaji and M. Imada, Phys. Rev. B {\bf 83}, 205122 (2011).
\bibitem{kmh4} D. Zheng, G.-M. Zhang, and C. Wu, Phys. Rev. B {\bf 84}, 205121 (2011).
\bibitem{kmh5} D.-H. Lee, Phys. Rev. Lett. {\bf 107}, 166806 (2011).
\bibitem{kmh6} S.-L. Yu, X. C. Xie, and J.-X. Li, Phys. Rev. Lett. {\bf 107}, 010401 (2011).
\bibitem{kmh7} M. Mardani, M.-S. Vaezi, and A. Vaezi, arXiv:1111.5980.
\bibitem{kmh8} J. Wen, M. Kargarian, A. Vaezi, and G. A. Fiete, Phys. Rev. B {\bf 84}, 235149 (2011).
\bibitem{kmh9} M. Hohenadler, T. C. Lang, and F. F. Assaad, Phys. Rev. Lett. {\bf 106}, 100403 (2011).
\bibitem{kmh10} C. Griset and C. Xu, Phys. Rev. B {\bf 85}, 045123 (2012).
\bibitem{kmh11} W. Wu, S. Rachel, W.-M. Liu, and K. Le Hur, Phys. Rev. B {\bf 85}, 205102 (2012).
\bibitem{kmh12} M. Hohenadler, Z. Y. Meng, T. C. Lang, S. Wessel, A. Muramatsu, and F. F. Assaad, Phys. Rev. B {\bf 85}, 115132 (2012).
\bibitem{kmh13} S. Ueda, N. Kawakami, and M. Sigrist, Phys. Rev. B {\bf 87}, 161108 (2013).
\bibitem{kmh14} H.-H. Hung, L. Wang, Z.-C. Gu, and G. A. Fiete, Phys. Rev. B {\bf 87}, 121113 (2013).
\bibitem{kmh15} H.-H. Hung, V. Chua, L. Wang, and G. A. Fiete, Phys. Rev. B {\bf 89}, 235104 (2014).
\bibitem{kmh16} Z. Y. Meng, H.-H. Hung, and T. C. Lang,  Mod. Phys. Lett B {\bf 28} 143001 (2014).
\bibitem{kmh17} Y. Araki, T. Kimura, A. Sekine, K. Nomura, and T. Z. Nakano, arXiv:1311.3973.
\bibitem{kmh18} Y. Araki and T. Kimura, Phys. Rev. B {\bf 87}, 205440 (2013).
\bibitem{kmh19} F. F. Assaad, M. Bercx, and M. Hohenadler, Phys. Rev. X {\bf 3}, 011015 (2013).
\bibitem{kmh20} M. Laubach, J. Reuther, R. Thomale, and S. Rachel, arXiv:1312.2934.
\bibitem{zare} M. H. Zare, F. Fazileh, and F. Shahbazi, Phys. Rev. B {\bf 87}, 224416 (2013).
\bibitem{vaezi} A.Vaezi, M.Mashkoori, and M. Hosseini, Phys. Rev. B {\bf 85}, 195126 (2012).
\bibitem{emrrvb} S. M. Bhattacharjee, Z. Phys. B: Condensed Matter, {\bf 82} 323 (1991).
\bibitem{mambrini} M. Mambrini, A. L\"{a}uchli, D. Poilbanc, and F. Mila, Phys. Rev. B, \textbf{74}, 144422 (2006)
\bibitem{chubukov} A. V. Chubukov and Th. Jolicoeur, Phys. Rev. B {\bf 44}, 12050 (1991).
\bibitem{sachdev} S. Sachdev and R. N. Bhatt, Phys. Rev. B {\bf 41}, 9323 (1990).
\end{thebibliography}
\end{document}